\documentclass[aps,prl,twocolumn]{revtex4-1}

\usepackage{graphicx}

\def\e0{\varepsilon_0}

\def\ddr{\partial_r}

\def\zv{\hat{\bf z}}
\def\rv{\hat{\bf r}}

\def\bea{\begin{eqnarray}}
\def\eea{\end{eqnarray}}

\begin{document}

\title{Toy model of the ``fountain effect''\\for magnetic field generation in intense laser-solid interactions}

\author{Andrea Macchi}\email{andrea.macchi@ino.it}
\affiliation{Istituto Nazionale di Ottica, CNR, research unit ``Adriano Gozzini'', Pisa, Italy}
\affiliation{Dipartimento di Fisica ``E. Fermi'', Largo B. Pontecorvo 3, 56127 Pisa, Italy}

\date{\today}

\begin{abstract}
A very simple ``geometrical'' model of a fountain is analyzed to evaluate the 
net flow on the ground resulting from the superposition of the source and the 
falling streams. 
On this basis we suggest a scaling for the magnetic field generated 
at the rear surface of high-intensity laser-irradiated solid targets due to the
``fountain effect'' related to fast electrons escaping in vacuum.
\end{abstract}

\maketitle

The interaction of superintense laser pulses with matter leads to the production
of a large number of energetic electrons and thus of very high electric currents
which, in turn, may lead to very strong quasi-static magnetic fields.
A particular case of interest is the interaction with thin solid targets, 
which is very relevant to schemes for ion acceleration 
\cite{borghesiFST06,*macchiRMP12}.
In such configuration the flow of ``fast'' electrons generated at front side 
of the target, i.e. at the laser-plasma interaction surface, may cross the
target and try to escape in vacuum. There, the electron flow can not be 
neutralized by a counterstreaming ``return'' current and thus electrons are 
stopped by the self-generated electric field. If the flow of electrons has 
a finite divergence, electrons may fall back to the target along curved 
trajectories, so that current loops may be generated along with a magnetic 
field. A similar ``fountain effect'' has been invoked to explain both 
experimental and numerical observations of magnetic fields for various 
laser-plasma conditions \cite{kolodnerPRL79,*pukhovPRL01}. 
However, to our knowledge the sound
qualitative description of the generation of magnetic fields by such effect
has been not supported by an analytical modeling capable to give estimates and 
scaling laws for the magnetic field. In the attempt to support the 
interpretation of recent measurements suggesting the generation of nearly 
hundreds of MGauss fields at the rear side of a solid target irradiated
at $10^{19}~\mbox{W cm}^{-2}$ \cite{sarriXXX12}, we formulated a very simple 
``geometrical'' model of a fountain to infer the dependence of the magnetic
field on the intensity, divergence and spatial extension of the electron flow.

\begin{figure}
\includegraphics[width=0.48\textwidth]{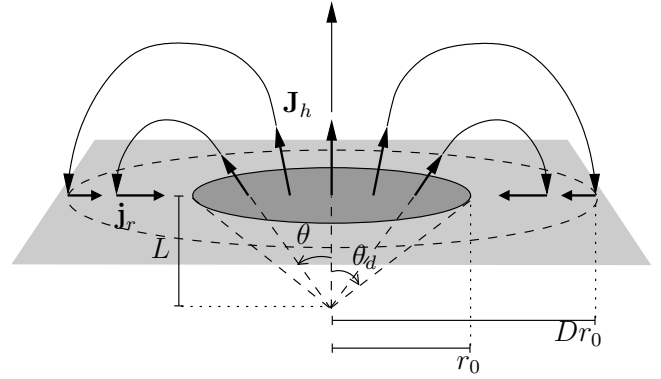}
\caption{Sketch of the ``toy'' fountain model.}
\label{fig:toyfountain}
\end{figure}

The ``toy'' model we use is sketched in 
Fig.\ref{fig:toyfountain}. The fountain is defined by a given 
configuration of the stream of ``water'' from the ground plane ($z=0$). 
The water ejected from the source falls back on the ground under the action of 
a gravity field ${\bf g}=-g\zv$. At any point on the ground, the total
flow is the difference of the source flow and of the flow of falling water.
To keep things simple, in our toy fountain the current source is defined at 
the plane $z=0$ by the distribution
\bea
{\bf J}_{\uparrow}(r)=J_0\left(1-\frac{r^2}{r_0^2}\right)\Theta(r_0-r)
(\zv\cos\theta+\rv\sin\theta) \, .
\eea
This means that the stream lines originate from a virtual source placed at a 
distance $L=r_0/\tan\theta_d$ under the $z=0$ surface, where $\theta_d$ is
the aperture angle (or divergence) of the fountain and $r_0$ is the radius
of the source area. In the limit $L=r_0=0$ the source is point-like and 
located on the ground. 

We assume the water particles to be ejected all with the same kinetic energy,
i.e. with the same modulus of the velocity $v_0$. Thus, at the radial position
$r=L/\tan\theta$, the velocity of the flow is 
${\bf v}_0(r)=(v_{0z},v_{0r})=v_0(\zv\cos\theta+\rv\sin\theta)$
with $(\cos\theta,\sin\theta)=(L,r)/(r^2+L^2)$.
Under the action of gravity these particles will fall back on the surface
after a time $t_f=2v_{0z}/g$ and at a distance $\Delta r=v_{0r}t_f$ from the
initial position. Then, the ``landing'' radius $r'$ is given as a function
of $r$ by
\bea
r'=r+\frac{2v_0^2}{g}\frac{rL}{r^2+L^2} \, .
\eea  
This equation defines the map from the source point to the falling point. 
It requires the divergence angle to be $\theta_d<\pi/4$, the angle for which 
$\Delta_r$ is maximum, otherwise the map would be singular. In the following we
restrict ourselves to the case of small divergence angles and thus we do not
care about the possible singularity.

\begin{figure}
\includegraphics[width=0.48\textwidth]{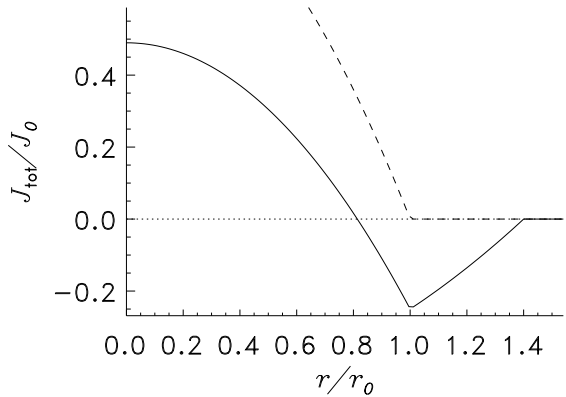}
\caption{Radial profiles of the total current (thick line) and the source 
current (dashed) for $D=1.4$.}
\label{fig:fountaincurrent} 
\end{figure}

The falling back of the water as ``rain'' is described by a current 
${\bf J}_{\downarrow}$ at the surface.
Due to the divergence of the flow ${\bf J}_{\downarrow}$ 
will not be exactly in the direction opposite to ${\bf J}_{\uparrow}$,
and will spread over a radius larger than $r_0$. 
Conservation of the flow implies
\bea
2\pi r'J_{\downarrow}(r')=2\pi rJ_{\uparrow}(r) \, , \\ 
J_{\downarrow}(r)=J_{\downarrow}[r'(r)]
=J_{\uparrow}(r)\frac{r}{r'}\frac{dr}{dr'} \, .
\eea
In order to keep the calculation simple and to highlight the main effects,
let us assume $r^2<r_0^2 \ll L^2$, i.e. $\theta_d\ll 1$, thus
\bea
r' \simeq r\left(1+\frac{2v_0^2}{gL}\right) \equiv Dr \, , \qquad (D>1) \, , 
\eea
so that the ``rain'' current spreads over a radius $D r_0>r_0$ and is given by
\bea
J_{\downarrow}(r)=\frac{J_0}{D^2}\left(1-\frac{r^2}{D^2r_0^2}\right)
                   \Theta\left(1-\frac{r}{Dr_0}\right) \, .
\eea
The total flow at the surface is thus
\bea
&J_{\mbox{\tiny tot}}(r) =J_{\uparrow}(r)-J_{\downarrow}(r) \nonumber \\
& =J_0\left\{\begin{array}{ll}
\left(1-\frac{1}{D^{2}}\right)\left(1-\frac{r^2}{r_0^2}\frac{1+D^2}{D^2}\right) & r<r_0 \\
-\frac{1}{D^{2}}\left(1-\frac{r^2}{(Dr_0)^2}\right) & r_0<r<Dr_0
\end{array}\right. \, .
\eea
The total current vanishes at $r=r_b \equiv r_0D^2/(1+D^2)<r_0$ and has 
a derivative cusp at $r=r_0$. 
Figure~\ref{fig:fountaincurrent} shows the profiles 
of the total current (thick line) and the source current (dashed) for 
$D=1.4$.
The flow may be balanced by a surface ``return'' current $j_r(r)$
flowing on the ground and determined by the equation 
$r^{-1}\ddr(rj_r)=J_{\mbox{\tiny tot}}(r)$.

Now suppose this toy fountain to provide a model for the distribution of the 
electric current due to fast 
electrons streaming out of the target (since the electron charge is negative,
the electric current will be actually in the direction opposite to the flow;
to keep track of the sign of $B_{\phi}$ in the following it is sufficient to
assume $J_0<0$).
Such current distribution will generate, just outside the surface
$z=0$, a magnetic field
\bea
B_{\phi}(r,z=0^{+})=\mu_0j_r(r) \, ,
\eea
within the assumption that $B_{\phi}(r,z)$ vanishes inside the target, i.e.
for $z<0$, which is reasonable if the target is conducting.  
The peak field will be at the point $r=r_b$. By integrating the above
equation from 0 to $r_b$
\bea
r_b B_{\phi}(r_b,z=^{+})&=&\mu_0 J_0(1-D^{-2})\int_0^{r_b}r(1-r^2/r_b^2)dr 
\nonumber \\
            &=&\mu_0J_0(1-D^{-2})r_b^2/4 \, .
\eea
In terms of the total current $I_0=J_0\pi r_0^2/2$ we thus write
\bea
B^{\mbox{\tiny (max)}}=B_{\phi}(r_b,z=0^{+})
=\frac{\mu_0I_0}{2\pi r_0}\frac{D^2-1}{D^2+1} \, .
\eea
Since we assumed $\theta_d \ll 1$, we may approximate
\bea
D=1+\frac{2v_0^2}{g r_0}\tan\theta_d \simeq 1+\frac{2v_0^2}{g r_0}\theta_d \, ,
\eea
and so if ${2v_0^2}/{g r_0}$ is not large with respect to unity, 
$D^2 \simeq 1+4v_0^2\theta_d/gr_0$ and
\bea
B^{\mbox{\tiny (max)}}\simeq
\frac{4v_0^2}{g r_0}\theta_d\frac{\mu_0I_0}{2\pi r_0} \, .
\eea
Thus the model roughly predicts that for small divergence the magnetic
field will scale as $\sim \theta_d B_0$ where 
$B_0={\mu_0I_0}/{2\pi r_0}$ is the peak field generated by a total current
$I_0$ distributed over a circle of radius $r_0$.
In addition, we notice that $z_0=v_0^2/2g$ is
the height of the fountain, i.e. the point along $z$ at which the water flow 
emitted in the perpendicular direction ($r=0$) gets to. Thus, we may also write
\bea
B^{\mbox{\tiny (max)}}\simeq
\frac{8z_0}{r_0}\theta_d\frac{\mu_0I_0}{2\pi r_0} \, .
\label{eq:Bgeom}
\eea
For the electron sheath at the rear of laser-irradiated solid targets the ratio
$z_0/r_0$ may be estimated in experiments by proton imaging data
\cite{romagnaniPRL05}, or directly in PIC simulations. At early times, before
the ions move and significant expansion occurs, typically $z_0\ll r_0$.

The above relation links $B^{\mbox{\tiny (max)}}$ to the geometrical ratio
$z_0/r_0$ and allows to test some consistency relations between the magnetic
field and the parameters of the electron flow. We may also try to relate to
evaluate the ``gravity field'' $g$ in terms of such parameters.
The gravity effect should be primarily provided by the back-holding electric
field, which of course is far from being uniform. 
Nevertheless, just to infer a possible scaling and an order of magnitude, 
let us estimate $g \simeq eE/m_e$ where $E$ is the typical value of the 
electric field near to the surface, whose value is related to the amount of
charge that leaves the target \cite{quinnPRL09}
and also to the energy of impurity protons accelerated from the surface
\cite{macchiRMP12}.
In addition, we substitute $m_e v_0^2/2=T_h$ where $T_h$ is the temperature of
hot electrons, a characterizing parameter of experiments in this context. 
Thus the magnetic field may be also estimated as
\bea
B^{\mbox{\tiny (max)}}\simeq
\frac{8T_h}{eE r_0}\theta_d\frac{\mu_0I_0}{2\pi r_0} \, .
\label{eq:Belec}
\eea

In the experiment of Ref.\cite{sarriXXX12}, evidence of magnetic fields at the 
rear surface having toroidal symmetry and a peak value of 
$\simeq 0.9\times 10^4~\mbox{T}$ has been provided by proton imaging data.
In the investigated experimental conditions, it has been estimated that a
laser pulse of $\simeq 10^{19}~\mbox{W cm}^{-2}$ intensity focused on a
spot area of $\simeq 10~\mu\mbox{m}$ radius converts a fraction
$f \simeq 0.1$ of its energy into ``fast'' electrons of $\simeq 0.5~\mbox{MeV}$ 
kinetic energy, resulting in a total electric current through the target 
$I_0=5 \times 10^6~\mbox{A}$. A beam divergence of 
$25^{\circ}=0.44~\mbox{rad}$
and a corresponding emitting area of radius $r_0 \simeq 15~\mu\mbox{m}$ have 
been estimated, so that $B_0 \simeq 6.7 \times 10^4~\mbox{T}$.
With these parameters, and roughly assuming ${z_0}/{r_0}\simeq 0.1$ as suggested
by PIC simulations performed in similar conditions, according to 
Eq.(\ref{eq:Bgeom}) our toy model predicts 
$B^{\mbox{\tiny (max)}} \simeq 8 \times 0.1 \times 0.44 B_0 \simeq 0.35B_0
=2.3 \times 10^{4}~\mbox{T}$, which is not far from the experimental results.
Moreover, in similar conditions 
the electric field at the surface $E \simeq 10^{12}~\mbox{V m}^{-1}$
and,  using this value, Eq.(\ref{eq:Belec})
gives $B^{\mbox{\tiny (max)}} \simeq 0.1B_0=0.7\times 10^4~\mbox{T}$,
which is also fairly consistent with the data.

Of course our formulas are expected to provide at best a correct order
of magnitude for the magnetic field, due to the extreme simplicity of the
toy model. Moreover, several assumptions are either weak or questionable 
such as e.g. non-relativistic electrons, small beam divergence, uniform 
electric field, and so on. 
In addition, the magnetic field is large enough for electrons to be
self-magnetized, so the electron cloud might be in (E-)MHD conditions and
its dynamics should be evaluated self-consistently, which is a difficult
task. Finally, the model may be appropriate (if at all) for early times
only, e.g. during the laser pulse (so that there is a continuous flow of 
electrons from the target), while at lower times convection and dissipative
effects would lead the field to decay \cite{sarriXXX12}.
That said, the toy model might provide a physical insight into the mechanism
of magnetic field generation and a scaling with ``geometrical''
parameters such as beam divergence and electron sheath extension.

\acknowledgments
Discussions with G. Sarri, M. Borghesi and F. Pegoraro and
support by MIUR, Italy through the FIRB project ``SULDIS'' are acknowledged.


\begin{thebibliography}{1}%
\makeatletter
\providecommand \@ifxundefined [1]{%
 \ifx #1\undefined \expandafter \@firstoftwo
 \else \expandafter \@secondoftwo
\fi
}%
\providecommand \@ifnum [1]{%
 \ifnum #1\expandafter \@firstoftwo
 \else \expandafter \@secondoftwo
\fi
}%
\providecommand \enquote [1]{``#1''}%
\providecommand \bibnamefont  [1]{#1}%
\providecommand \bibfnamefont [1]{#1}%
\providecommand \citenamefont [1]{#1}%
\providecommand\href[0]{\@sanitize\@href}%
\providecommand\@href[1]{\endgroup\@@startlink{#1}\endgroup\@@href}%
\providecommand\@@href[1]{#1\@@endlink}%
\providecommand \@sanitize [0]{\begingroup\catcode`\&12\catcode`\#12\relax}%
\@ifxundefined \pdfoutput {\@firstoftwo}{%
 \@ifnum{\z@=\pdfoutput}{\@firstoftwo}{\@secondoftwo}%
}{%
 \providecommand\@@startlink[1]{\leavevmode}%
 \providecommand\@@endlink[0]{}%
}{%
 \providecommand\@@startlink[1]{%
  \leavevmode
  \pdfstartlink
   attr{/Border[0 0 1 ]/H/I/C[0 1 1]}%
   user{/Subtype/Link/A<</Type/Action/S/URI/URI(#1)>>}%
  \relax
 }%
 \providecommand\@@endlink[0]{\pdfendlink}%
}%
\providecommand \url  [0]{\begingroup\@sanitize \@url }%
\providecommand \@url [1]{\endgroup\@href {#1}{\urlprefix}}%
\providecommand \urlprefix [0]{URL }%
\providecommand \Eprint[0]{\href }%
\@ifxundefined \urlstyle {%
  \providecommand \doi [1]{doi:\discretionary{}{}{}#1}%
}{%
  \providecommand \doi [0]{doi:\discretionary{}{}{}\begingroup
  \urlstyle{rm}\Url }%
}%
\providecommand \doibase [0]{http://dx.doi.org/}%
\providecommand \Doi[1]{\href{\doibase#1}}%
\providecommand \bibAnnote [3]{%
  \BibitemShut{#1}%
  \begin{quotation}\noindent
    \textsc{Key:}\ #2\\\textsc{Annotation:}\ #3%
  \end{quotation}%
}%
\providecommand \bibAnnoteFile [2]{%
  \IfFileExists{#2}{\bibAnnote {#1} {#2} {\input{#2}}}{}%
}%
\providecommand \typeout [0]{\immediate \write \m@ne }%
\providecommand \selectlanguage [0]{\@gobble}%
\providecommand \bibinfo [0]{\@secondoftwo}%
\providecommand \bibfield [0]{\@secondoftwo}%
\providecommand \translation [1]{[#1]}%
\providecommand \BibitemOpen[0]{}%
\providecommand \bibitemStop [0]{}%
\providecommand \bibitemNoStop [0]{.\EOS\space}%
\providecommand \EOS [0]{\spacefactor3000\relax}%
\providecommand \BibitemShut [1]{\csname bibitem#1\endcsname}%
\bibitem{borghesiFST06}%
  \BibitemOpen
  \bibfield{author}{%
  \bibinfo {author} {\bibfnamefont{M.}~\bibnamefont{Borghesi}} \emph{et~al.},\
  }%
  \bibfield{journal}{%
  \bibinfo {journal} {Fus. Sci. Techn.}\ }%
  \textbf{\bibinfo {volume} {49}},\ \bibinfo {pages} {412} (\bibinfo {year}
  {2006})%
  \bibAnnoteFile{NoStop}{borghesiFST06}%
\bibitem{macchiRMP12}%
  \BibitemOpen
  \bibfield{author}{%
  \bibinfo {author} {\bibfnamefont{A.}~\bibnamefont{Macchi}}, \bibinfo {author}
  {\bibfnamefont{M.}~\bibnamefont{Borghesi}},\ and\ \bibinfo {author}
  {\bibfnamefont{M.}~\bibnamefont{Passoni}},\ }%
  \bibfield{journal}{%
  \bibinfo {journal} {Rev. Mod. Phys.}}%
   (\bibinfo {year} {2012}),\ \bibinfo {note} {to be published}%
  \bibAnnoteFile{NoStop}{macchiRMP12}%
\bibitem{kolodnerPRL79}%
  \BibitemOpen
  \bibfield{author}{%
  \bibinfo {author} {\bibfnamefont{P.}~\bibnamefont{Kolodner}}\ and\ \bibinfo
  {author} {\bibfnamefont{E.}~\bibnamefont{Yablonovitch}},\ }%
  \bibfield{journal}{%
  \Doi{10.1103/PhysRevLett.43.1402}{\bibinfo {journal} {Phys. Rev. Lett.}}\ }%
  \textbf{\bibinfo {volume} {43}},\ \bibinfo {pages} {1402} (\bibinfo {year}
  {1979})%
  \bibAnnoteFile{NoStop}{kolodnerPRL79}%
\bibitem{pukhovPRL01}%
  \BibitemOpen
  \bibfield{author}{%
  \bibinfo {author} {\bibfnamefont{A.}~\bibnamefont{Pukhov}},\ }%
  \bibfield{journal}{%
  \Doi{10.1103/PhysRevLett.86.3562}{\bibinfo {journal} {Phys. Rev. Lett.}}\ }%
  \textbf{\bibinfo {volume} {86}},\ \bibinfo {pages} {3562} (\bibinfo {year}
  {2001})%
  \bibAnnoteFile{NoStop}{pukhovPRL01}%
\bibitem{sarriXXX12}%
  \BibitemOpen
  \bibfield{author}{%
  \bibinfo {author} {\bibfnamefont{G.}~\bibnamefont{Sarri}} \emph{et~al.},\ }%
  \bibfield{journal}{%
  \bibinfo {journal} {Phys. Rev. Lett.}}%
   (\bibinfo {year} {2012}),\ \bibinfo {note} {submitted for publication}%
  \bibAnnoteFile{NoStop}{sarriXXX12}%
\bibitem{romagnaniPRL05}%
  \BibitemOpen
  \bibfield{author}{%
  \bibinfo {author} {\bibfnamefont{L.}~\bibnamefont{Romagnani}} \emph{et~al.},\
  }%
  \bibfield{journal}{%
  \Doi{10.1103/PhysRevLett.95.195001}{\bibinfo {journal} {Phys. Rev. Lett.}}\
  }%
  \textbf{\bibinfo {volume} {95}},\ \bibinfo {pages} {195001} (\bibinfo {year}
  {2005})%
  \bibAnnoteFile{NoStop}{romagnaniPRL05}%
\bibitem{quinnPRL09}%
  \BibitemOpen
  \bibfield{author}{%
  \bibinfo {author} {\bibfnamefont{K.}~\bibnamefont{Quinn}} \emph{et~al.},\ }%
  \bibfield{journal}{%
  \Doi{10.1103/PhysRevLett.102.194801}{\bibinfo {journal} {Phys. Rev. Lett.}}\
  }%
  \textbf{\bibinfo {volume} {102}},\ \bibinfo {pages} {194801} (\bibinfo {year}
  {2009})%
  \bibAnnoteFile{NoStop}{quinnPRL09}%
\end{thebibliography}
\end{document}